\documentclass[preprint,12pt]{elsarticle}

\usepackage{amssymb}
\begin{document}
\begin{frontmatter}

\title{Observation of lattice softening at $T_{\rm c}$ in the FeSe$_{0.5}$Te$_{0.5}$
superconductor
}

\author{J. Lind\'en\corref{cor1}}
\cortext[cor1]{Corresponding author. E-mail address: jlinden@abo.fi}
\author{J.-P. Lib\"ack}
\address{Department of Physics, \AA bo Akademi, FI-20500 Turku, Finland}
\author{M. Karppinen, E.-L. Rautama, H. Yamauchi}
\address{Department of Chemistry, Aalto University School of Science and Technology, FI-00076 Aalto, Finland}

\begin{abstract}
Polycrystalline samples of FeSe$_{0.5}$Te$_{0.5}$ were synthesized using a conventional solid-state reaction method.
The onset of bulk superconductivity transition was confirmed
by SQUID magnetometry at 12.5~K. $^{57}$Fe M\"ossbauer spectra in transmission geometry were recorded at
temperatures between 6.0 and 320 K. Both the isomer shift and the total absorption started to drop about $T_c$, indicating a softening of the lattice. The drop is estimated to correspond to at least 60~K from the original Debye temperature
$\theta_{\rm D}\approx 460$~K.
Seebeck measurements indicate that the samples are $n$-type conductors at low temperatures with a cross-over to $p$-type conductivity around 135 K. The zero Seebeck coefficient is seen below $10.6$~K.
\end{abstract}
\end{frontmatter}

%\pacs{76.80.+y, 74.70.Xa}

\section{INTRODUCTION}
\label{intro}
Superconductivity above 20 K in iron pnictides was reported two years ago.\cite{Hosono} Since then the number of iron-pnictide phases has grown considerably, but the highest critical temperature has remained below 60 K. However, the
value falls short only of that of the high-$T_c$ superconductive cuprates. Nonetheless, Fe as a natural constituent in
superconductive materials is fascinating and somewhat unexpected. In none of these iron-pnictide phases does iron order magnetically in the superconductive state, though some non-superconductive parent phases exhibit antiferromagnetic spin-density waves.\cite{SDW} Some of the superconductive iron phases have simple structures, e.g. the chalcogenide
phase FeSe.\cite{Hsu} Like the high-$T_{\rm c}$ cuprate superconductors Fe-based phases also have layered structures, in which the
superconducting charge carriers reside at edge-shared Fe$X_4$ ($X=$ Group 15 or 16 element) tetrahedra.
The geometry is important for both types of superconductive phases. Cuprate superconductors with puckered CuO$_2$ layers have generally lower $T_{\rm c}$ values than those with flat ones, whereas
for the Fe-based superconductors the highest $T_{\rm c}$ values occur when the tetrahedra are closest to the regular shape.\cite{Lee} In a
recent work which incorporates also FeSe the distance between the $X$ atom and the nearest Fe layer seems to be
a crucial geometrical parameter for determining the $T_{\rm c}$ value.\cite{Okabe}
However, the actual mechanism behind the superconductivity transition is not known yet. 

The presence of Fe has opened the possibility 
to perform $^{57}$Fe-M\"ossbauer-spectroscopy measurements and thereby to determine the spin and valence state of the Fe atoms. Various spectra from superconductive samples have established that the spin state of Fe is zero or close to zero.\cite{Stadnik,Kitao} This is true
at least within the characteristic time-scale of $\sim 100$~ns of the $^{57}$Fe transition. However, presence of antiferromagnetic
fluctuations has been suggested\cite{Christianson, Ikubo} but may be difficult to be detected by M\"ossbauer spectroscopy.
A recent high-pressure experiment on the LaFeAsO phase showed that there is a strong connection between the spin-density-wave type magnetism
of the non-superconducting state and the seemingly zero-Fe-spin spectra of the superconducting state.\cite{Kawakami}

In this work we report findings of the M\"ossbauer and Seebeck measurements performed on both sides of
the critical temperature of FeSe$_{0.5}$Te$_{0.5}$ to record possible changes in the hyperfine parameters upon passing $T_c$.   

\section{Experimental} 
\label{exp}
Samples of the FeSe$_{0.5}$Te$_{0.5}$ phase were synthesized from stoichiometric ratios of
high-purity Se (99.99\%), Fe (99.9\%) and  Te (99.8\%) powders. 
The powders
were sealed under vacuum in a quartz tube and fired for 20 h at 600 $^\circ$C. After an intermediate
grinding the samples were fired again at 650 $^\circ$C for 20 h and furnace-cooled down to 300 K.

Bulk superconductivity was confirmed by measuring the diamagnetic signal from approximately 20 mg of the sample with a superconducting
quantum interference device (SQUID: Quantum Design, MPMS-XL) in both the zero field-cooled and field-cooled (10 Oe) regimes. 

X-ray diffraction patterns were measured in an ordinary $\theta-2\theta$ geometry using Cu $K\alpha 1$ radiation (PanAnalytical X'Pert Pro MPD diffractometer) to check the sample purity. Rietveld analysis was done to confirm the tetragonal P4/nmm structure of superconductive $\beta$-FeSe phase.

The M\"ossbauer measurements were carried out
using a 25~mCi $^{57}$Co:$Rh$  source (Cyclotron Co, purchased in April 2009) 
at fixed temperatures between 6.0 and 320~K, in transmission geometry with a maximum Doppler velocity of 3.0~mm/s. Additional measurements using maximum Doppler velocities of 8 and 10~mm/s were done to check the
presence of magnetic impurities. Cooling of the sample was achieved with an Oxford CF506 continuous-flow
cryostat with liq. N$_2$ and liq. He as cooling agents. With the amount of He allocated for the measurements spectra from 6.0 to
30 K could be collected, leaving the interval of 30 - 77 K uncovered.
The two main components of the spectra were fitted using 
the chemical isomer shift relative to $\alpha$-Fe ($\delta$), the relative intensity ($I$),
the quadrupole coupling constant ($eQV_{zz}$), and the resonance line width ($\Gamma$) which 
was constrained to be equal for all components. 
A weak magnetic component due to magnetite was observed, but was left unfitted in most spectra
as the most intensive lines were outside the $\pm 3$-mm/s velocity range. 

Seebeck measurements were done (using a self-made equipment) upon heating the sample from 4.0 to 300 K in steps
of $\sim 0.3$~K.
A sintered sample specimen was ground into shape of a thin parallelepiped of $2.0\times 1.5\times 0.3$~mm$^3$ and inserted between two Cu sample holders using Ag paste. A resistor hidden inside one of the Cu holders was heated to
achieve a temperature gradient up to $\sim 1.5$~K over the sample. 
Additional Seebeck data in applied fields of 0 and 8 T were collected using a physical-property measurement system (Quantum Design, PPMS) equipped with a self-made sample puck. The same sample as above was used.

\section{Results and Discussion}
\begin{figure}
\includegraphics[width=\linewidth]{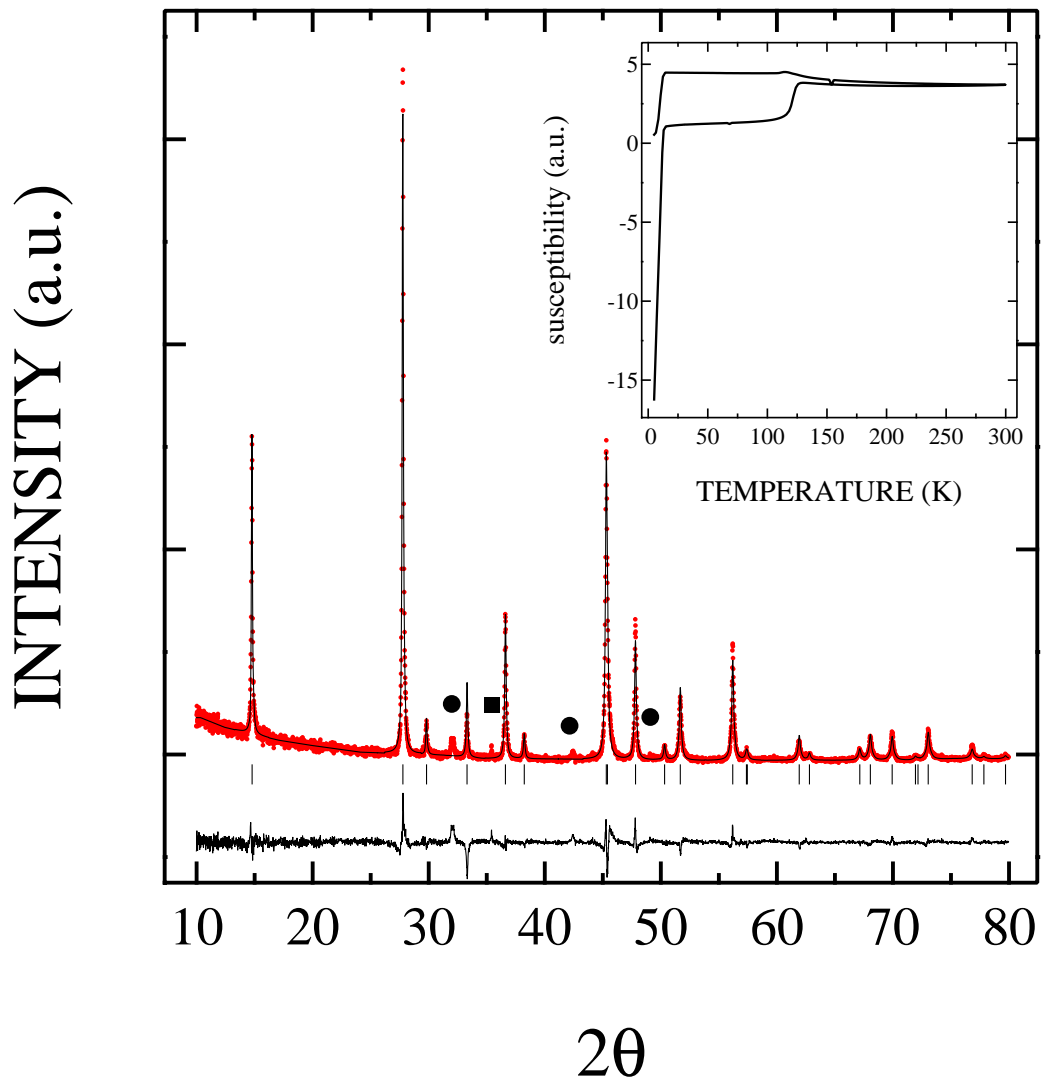}
\vspace{-3 mm}
\caption{(color online) X-ray diffraction pattern for the FeSe$_{0.5}$Te$_{0.5}$ sample. Impurity
peaks due to hexagonal Te-substituted $\delta-$FeSe/Fe$_7$Se$_8$ and Fe$_3$O$_4$ are indicated with filled circles and
squares, respectively. The inset shows the DC magnetic susceptibility measured in both zero-field-cooling and field-cooling regimes. 
}
\label{xrd}
\end{figure}

Figure~\ref{xrd} shows the  X-ray diffraction pattern for the FeSe$_{0.5}$Te$_{0.5}$ sample. It is readily indexed by the PbO-type tetragonal structure of space group P4/nmm (isostructural with $\beta-$FeSe). 
Impurity peaks assigned mainly to hexagonal Te-substituted $\delta$-FeSe\cite{Stadnik} and/or Fe$_7$Se$_8$ are observed. Additionally, presence of Fe$_3$O$_4$, confirmed by SQUID and M\"ossbauer spectroscopy measurements, is seen.

The magnetization-versus-temperature data obtained by SQUID measurements in the field-cooled and zero-field-cooled regimes confirm onset of bulk superconductivity at 12.5~K for the
FeSe$_{0.5}$Te$_{0.5}$ sample, see the inset in Fig.~\ref{xrd}. The positive background is due to the presence
of magnetic impurity phases, i.e. Fe$_3$O$_4$. The Verwey transition\cite{Verwey} of Fe$_3$O$_4$ at $\sim 120$~K is 
also seen in the susceptibility data.

\begin{figure}
\includegraphics[width=\linewidth]{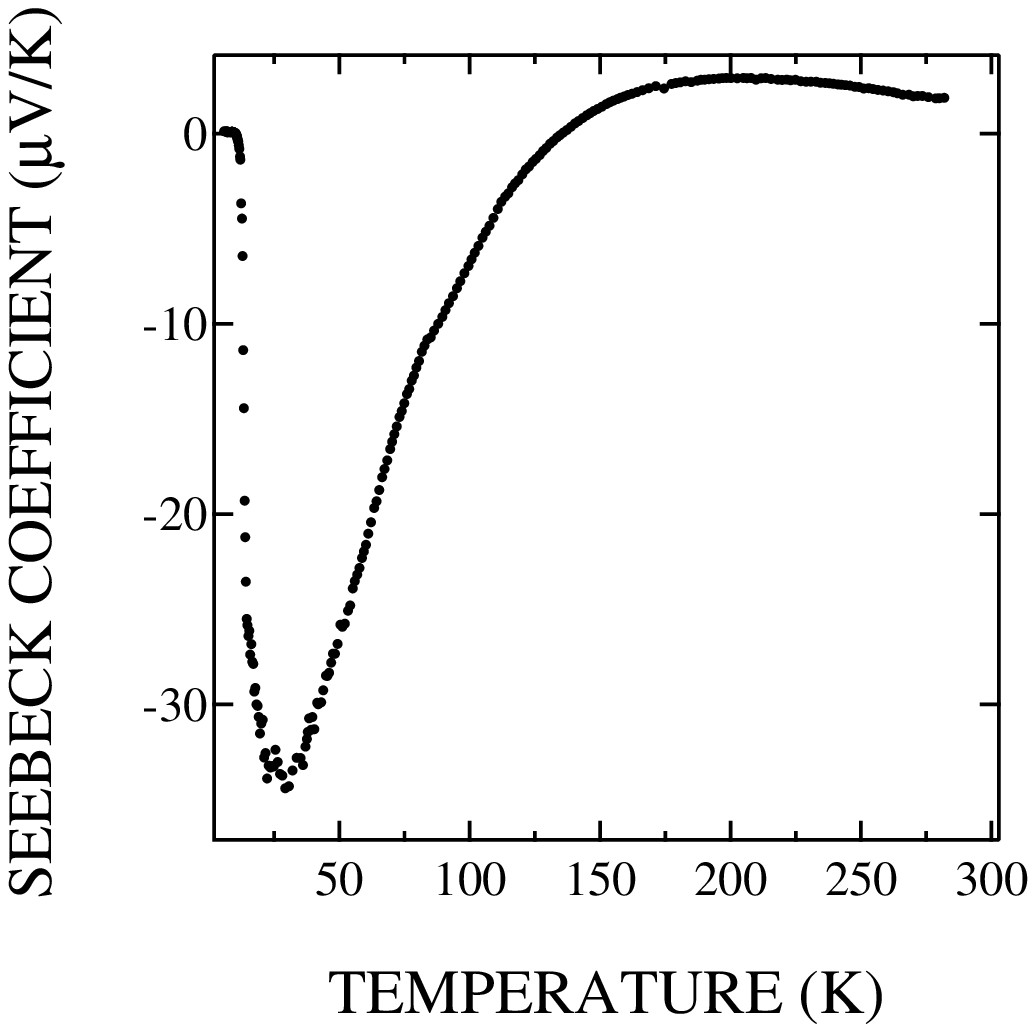}
\vspace{-3 mm}
\caption{Seebeck data for the FeSe$_{0.5}$Te$_{0.5}$ sample. 
}
\label{Seebeck}
\end{figure}

The Seebeck data are shown in Fig.~\ref{Seebeck}.
The negative sign indicates that the sample is an $n$-type conductor. Below the transition into the superconducting state
the Seebeck coefficient becomes zero at 10.6~K. Above $\sim 135$~K positive values for the Seebeck coefficient
are obtained. Using an external field of  8 T the low-temperature Seebeck data are clearly shifted by $\sim 1$~K towards lower
temperatures. The shift is in accord with the rather high second critical field $H_{{\rm c}_2}$.
Similar results were reported in Ref.~\cite{Pallecchi} for samples of Fe$_{1+x}$Te$_{1-y}$Se$_y$.

\begin{figure}
\includegraphics[width=9.6cm]{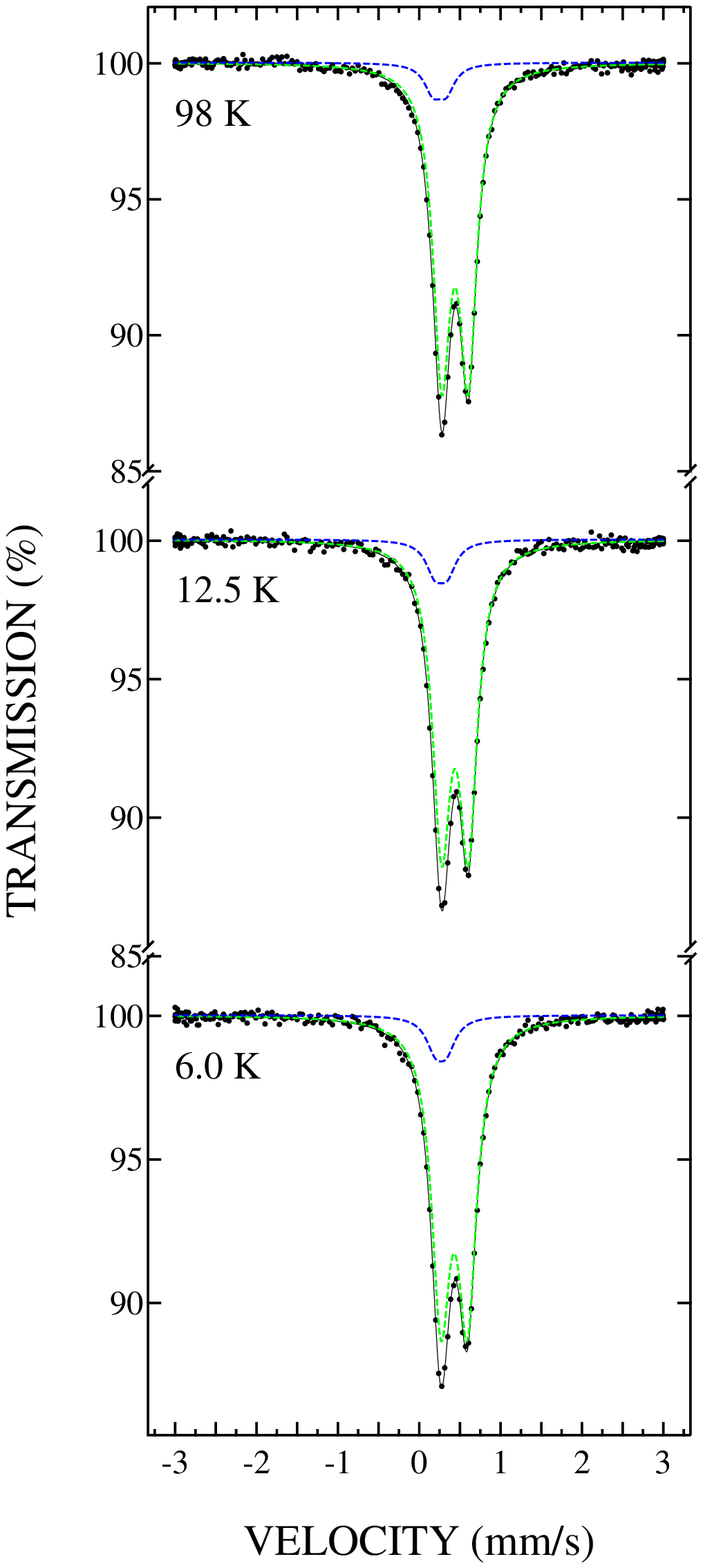}
\caption{(color online) M\"ossbauer spectra measured from the FeSe$_{0.5}$Te$_{0.5}$ sample at indicated temperatures. The components used in the fitting are displayed above each data set. 
}
\label{MossSeTe}
\end{figure}

M\"ossbauer spectra  obtained from FeSe$_{0.5}$Te$_{0.5}$ at selected temperatures are shown in Fig.~\ref{MossSeTe}.
Components due to two phases are readily observed. The paramagnetic main component covering 88\% of the spectral intensity
is due to FeSe$_{0.5}$Te$_{0.5}$ (Component 1). It has an isomer shift of $\sim 0.43$~mm/s at 300 K and
a quadrupole splitting of $eQV_{\rm zz}/2=0.28$~mm/s, i.e. similar to the values obtained for $\beta-$FeSe,\cite{Stadnik} but $eQV_{\rm zz}$ was somewhat larger, reflecting a higher deformation of the coordination tetrahedron around Fe upon introducing Te.\cite{Tegel}
Another paramagnetic component (Component 2) with $\delta\approx 0.36$~mm/s covers $\sim 9$\% of the spectral intensity.
It has a quadrupole splitting of 0.16~mm/s. In the XRD pattern for a $\beta-$FeSe sample
peak positions due to $\delta$-FeSe/Fe$_7$Se$_8$ impurities are observed.\cite{Stadnik} When comparing the pattern with that of the present FeSe$_{0.5}$Te$_{0.5}$ sample almost identical impurity peaks are observed, indicating that Component 2
is due to Te-substituted Fe$_7$Se$_8$ or $\delta$-FeSe. Fe$_7$Se$_8$ is magnetically ordered
at room-temperature, but the absence of magnetic order is in accord with the fact that Te substitution is detrimental for the magnetic order in the Fe-Se system.~\cite{FeSeTe_old_paper}
In a high-velocity spectrum (not shown) $\sim 4$\% of Fe$_3$O$_4$ can be discerned. No other magnetic phases were observed.

As Component 2 interferes with Component 1 we tried the following three fitting schemes when analyzing the spectra of the most
interesting region (6-100 K): (i) Releasing all fit parameters of Components 1 and 2, (ii) Fixing the peak height and quadrupole coupling constant of Component 2 to an average of the $T<100$~K region and (iii) Fixing the isomer shift and quadrupole coupling constant of Component 2 to an average value of the $T<100$~K region.
The general features as presented below for Component 1 did not depend on the choice of scheme. Fitting scheme (iii) gave the "purest" behavior, as no specific temperature evolution was expected for the impurity component, except that
its peak height should vary as a function of the temperature, which was violated by scheme (ii). In the end, we decided to present the results as obtained according to scheme (i), in order to avoid excessive fixing of fit parameters.

\begin{figure}
\includegraphics[width=\linewidth]{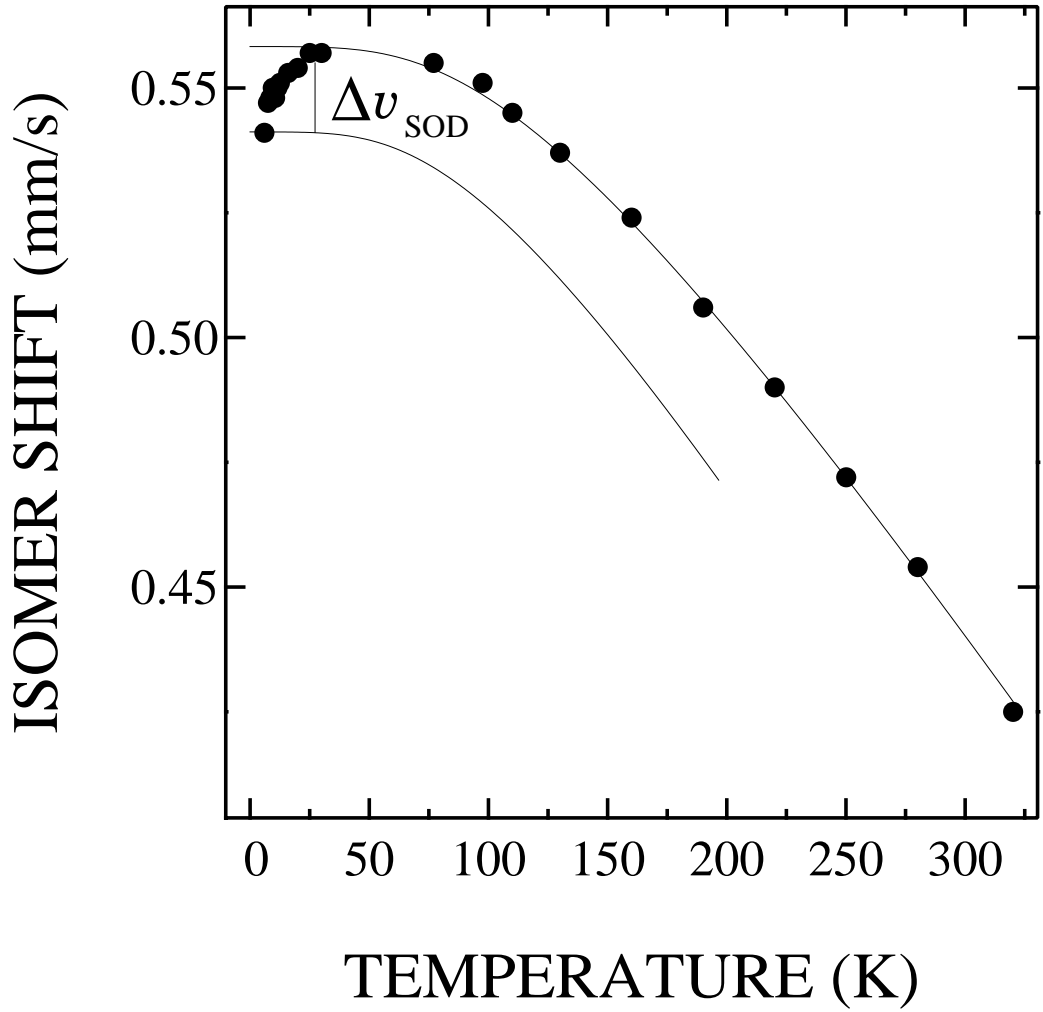}
\vspace{-3 mm}
\caption{Isomer shift vs. temperature for Component 1 of FeSe$_{0.5}$Te$_{0.5}$. The data was fitted using the full temperature
dependence for the second-order Doppler shift. A second curve corresponding to a Debye temperature of 400 K is drawn
for illustrating the size of the drop at $T_{\rm c}$.
}
\label{isomer_SeTe}
\end{figure}

The isomer shift ($\delta$) of Component 1 in FeSe$_{0.5}$Te$_{0.5}$ is plotted against temperature in Fig.~\ref{isomer_SeTe}. 
The shift obtained from the fitting includes contributions from 
both the chemical shift and the second-order Doppler shift, which is known to increase convexely 
upon decreasing temperature, due to gradual depopulation of the excited phonon states.\cite{Greenwood} However, it should be constant at low temperatures, because of the quantum mechanical zero-point motion. The chemical shift should not depend on temperature. An ordinary convex temperature evolution is indeed traced down to $T_c$ where a distinct drop is observed. The drop indicates that the Debye temperature ($\theta_{\rm D}$) is decreasing, that is the lattice undergoes a softening.
The actual size of the drop can only be estimated, as there is no leveling-off visible at 6 K.
The temperature dependence for $\delta$ is given by:\cite{Nasu}
\begin{equation}
\delta(T)=\delta_0+{{9k_{\rm B}\theta_{\rm D}}\over{16Mc}}-{{9k_{\rm B}T}\over{2Mc}}\left({T\over{\theta_{\rm D}}}\right)^3
\int_0^{{\theta_{\rm D}}\over T} {{x^3dx}\over{e^x-1}},
\label{isoT}
\end{equation}
where $c$ is the velocity of light,
$k_{\rm B}$ the Boltzmann constant, $M$ the mass of the $^{57}$Fe atom, and $\delta_0$ the temperature-independent part, i.e the chemical shift.
A fit with Eq. \ref{isoT} to the data of Fig.~\ref{isomer_SeTe} yields $\theta_{\rm D}=460(5)$~K 
and $\delta_0=0.43$~mm/s. By fixing $\delta_0$ at the values 0.43~mm/s and setting $\theta_{\rm D}=400$~K the
softening of the lattice is readily visualized: The drop in the second-order Doppler shift of at least $\Delta v_{\rm SOD}=-0.013$~mm/s corresponds to a drop of $\Delta \theta_{\rm D}=60$~K in the Debye temperature (the lower curve in Fig.~\ref{isomer_SeTe}).  

\begin{figure}
\includegraphics[width=\linewidth]{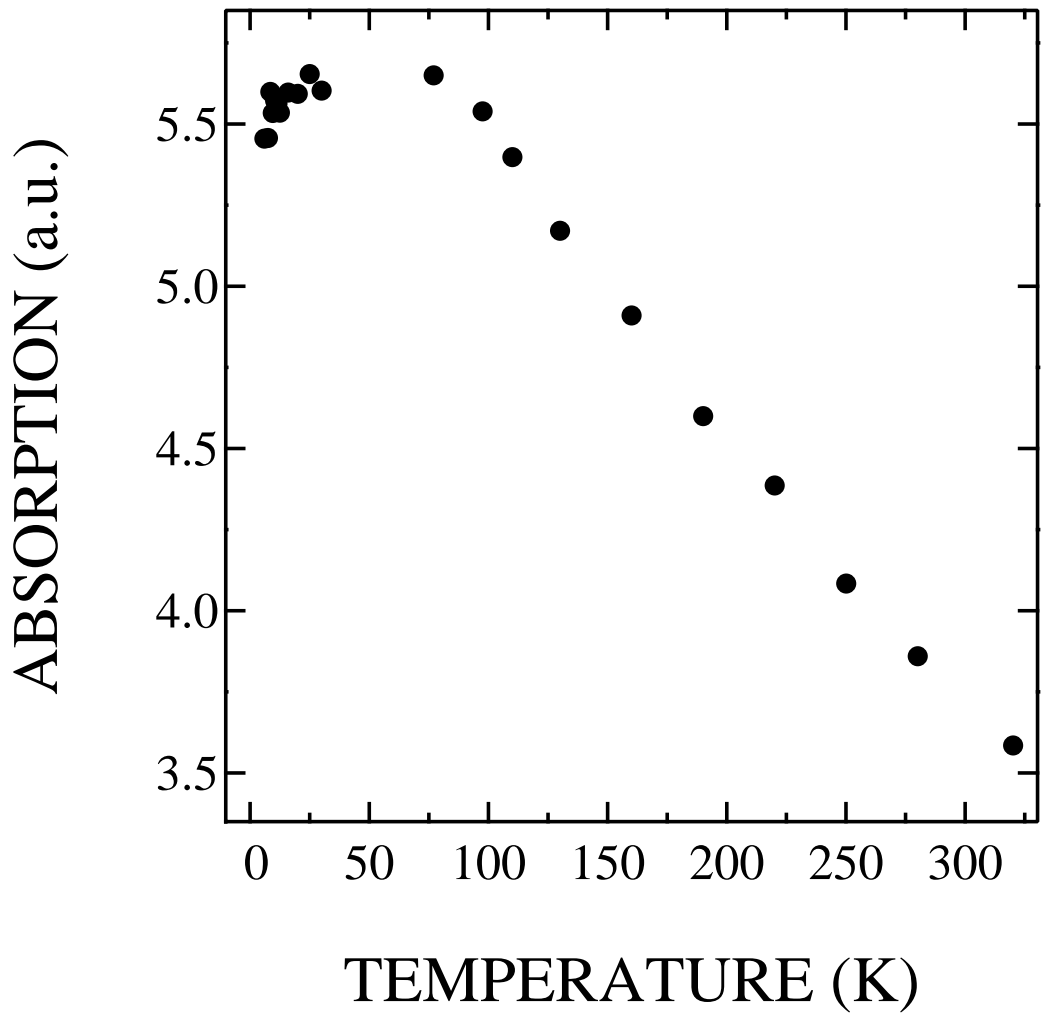}
\vspace{-3 mm}
\caption{M\"ossbauer absorption of Component 1 vs. temperature.
}
\label{FeSeTe_absorption}
\end{figure}

\begin{figure}
\includegraphics[width=\linewidth]{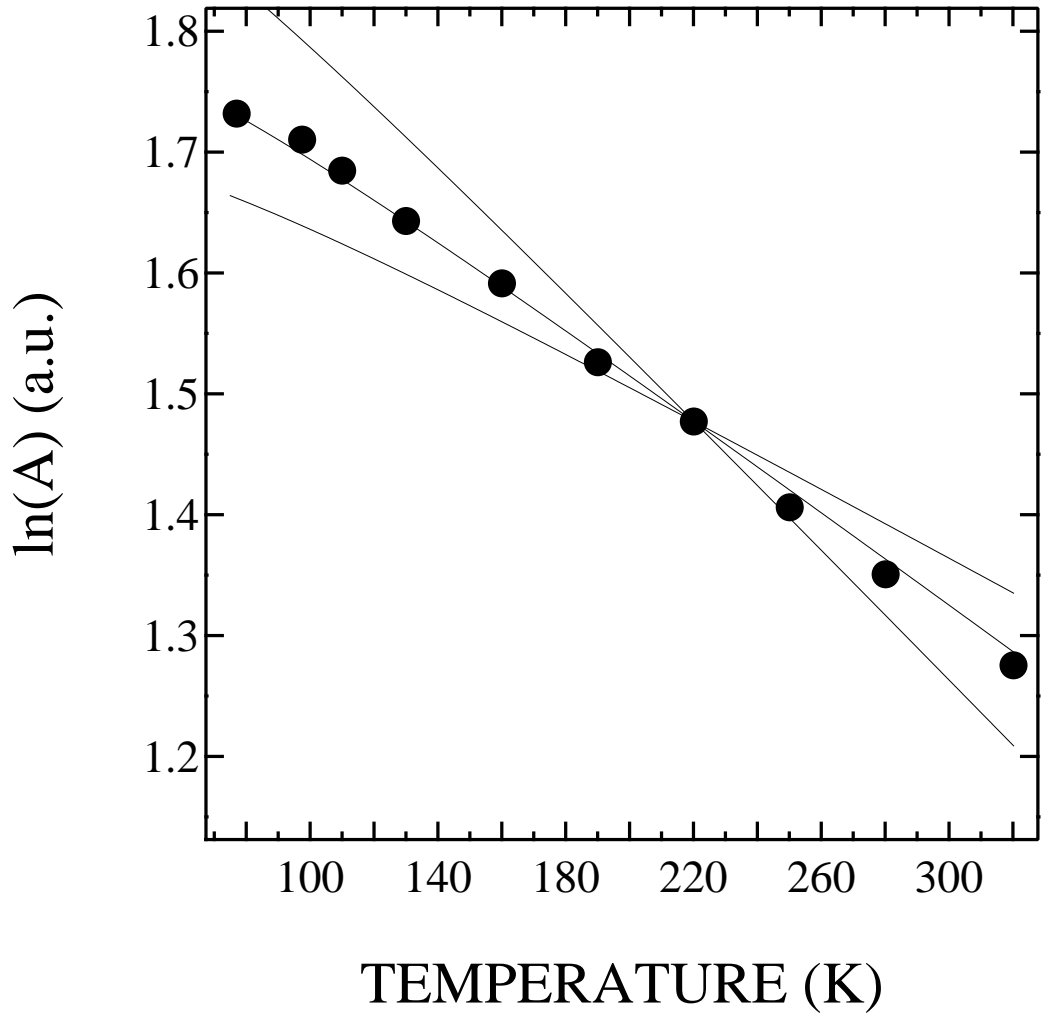}
\vspace{-3 mm}
\caption{Logarithm of absorption vs. temperature fitted using Eq.~\ref{lnf} with $\theta_{\rm D}=260$~K. Lines
corresponding to Debye temperatures of 220 and 300 K are also drawn.
}
\label{Fplot}
\end{figure}

The drop in  $\theta_{\rm D}$ is also visible in the total absorption ($A$) for Component 1, given as the component intensity $I$ times the line width $\Gamma$, as shown in Fig.~\ref{FeSeTe_absorption}. 
The drop in absorption begins slightly above $T_c$. A softening of the lattice
will decrease the recoil-free fraction and hence the absorption will also decrease.
A part of the leveling-off of the absorption could be due to the fact that the sample is rather thick, but that cannot explain the actual drop. Because of possible sample-thickness corrections
the absorption data can only be used for qualitatively estimating $\theta_{\rm D}$ and the drop at $T_{\rm c}$. The standard expression for the recoil-free fraction $f$ is given by:\cite{Green}
\begin{equation}
\ln f=-{{6E_R}\over{k_{\rm B}\theta_{\rm D}}}\left[{1\over 4}+\left({T\over{\theta_{\rm D}}}\right)^2
\int_0^{{\theta_{\rm D}}\over T} {{xdx}\over{e^x-1}}\right],
\label{f-factor}
\end{equation}
where the recoil energy $E_R=0.0019$~eV for the 14.4 keV $\gamma$ quantum of $^{57}$Fe. 
The absorption data of Fig.~\ref{FeSeTe_absorption} should be proportional
to the recoil-free fraction in the absorber, i.e. $A=Cf$, where $C$ is a parameter depending on the
sample thickness, signal-to-noise ratio for the detected $\gamma$ quanta etc. 
Thus:
\begin{equation} 
\ln A=\ln f-\ln C.
\label{lnf}
\end{equation}
Eq. \ref{lnf} is plotted together with the experimental $\ln(A)$ data against temperature in Fig.~\ref{Fplot}.
The data below 77 K where parameter $C$ is no longer constant are omitted in Fig~\ref{Fplot} because of the leveling-off of absorption data.
The slope of the curve strongly depends on the Debye temperature. The best fit was obtained for
$\theta_{\rm D}= 260(5)$~K, which is much lower than
the estimate based on the second-order Doppler shift. In fact, due to possible "hidden" temperature-dependence in parameter $C$,
the $\theta_{\rm D}=460$~K is likely to be more reliable. 

At low temperatures Eq.~\ref{f-factor} for the recoil-free fraction can be approximated with:
\begin{equation}
f\approx \exp\left[-{{3E_R}\over{2k_{\rm B}\theta_{\rm D}}}\right].
\label{fnaught}
\end{equation} 
Upon inserting the estimated $\theta_{\rm D}= 260$~K in Eq.~\ref{fnaught} the recoil-free
fraction before the drop is estimated  at $f=0.881$.
About $T_{\rm c}$ the absorption (shown in Fig.~\ref{FeSeTe_absorption}) drops from 5.63 to 5.45, which
corresponds to a drop in $f$ from 0.881 to 0.853. Upon inserting the latter into
Eq.~\ref{fnaught} we get $\theta_{\rm D}= 208$~K, resulting in a drop of $\sim 50$~K in $\theta_{\rm D}$. 
If the 460-K estimate is used the recoil-free factors before and after the drop are 0.93 and 0.90, respectively and
the corresponding drop in $\theta_{\rm D}$ is 140 K. 

The quadrupole splitting of Component 1  (not shown) increases from $eQV_{zz}/2=0.28$~mm/s to 0.33~mm/s as temperature is decreased below 100~K. A leveling-off at 0.33~mm/s, possible followed by a tiny drop of 0.01~mm/s around $T_{\rm c}$ is observed.
A small relaxation in the position of Fe or Se/Te atoms is all that is needed to cause a change in $eQV_{zz}$. The quantity $V_{zz}$  depends on the nearest-neighbor
distance ($r$) as $V_{zz}\sim 1/r^3$. That is, it is extremely sensitive to movements of the atoms. The movements need not to be commensurable, as the M\"ossbauer-resonant nuclei act as a local probes.

We have also synthesized samples of the $\beta$-FeSe phase and observed a parallel drop
in $\theta_{\rm D}$ about its $T_{\rm c}\approx 8.0$~K.\cite{ownFeSe} We recently reported
drops in both the isomer shift and the absorption below $T_{\rm c}=15$~K
for the superconducting
LiFeAs phase.\cite{Gao} In this phase the drops were preceded by a lattice stiffening, giving the low-temperature
absorption and isomer-shift curves concave shapes. 
Generally some change in the lattice stiffness, upon passing $T_{\rm c}$, might indicate that the BCS mechanism of 
phonon-mediated coupling between the superconducting charge carriers are active. However, the present lattice softening could
be the result of other mechanisms as well. In particular, antiferromagnetic fluctuations about $T_{\rm c}$ have been observed
in some of the iron-based pnictide and chalcogenide phases by inelastic neutron scattering, {\it cf.}  Refs.~\cite{Christianson, Wen}.
A possible connection between such fluctuations and the lattice softening was in fact suggested recently.\cite{Yildirim}
At the same time inelastic X-ray-scattering experiments point towards the importance of phonons in these phases.\cite{Chul-Ho,Reznik}   

\section{CONCLUSIONS}
\label{conclusions}
$^{57}$Fe M\"ossbauer measurements of the FeSe$_{0.5}$Te$_{0.5}$ phase reveal a paramagnetic quadrupole doublet with
an isomer shift compatible with zero spin Fe$^{2+}$. Upon decreasing the temperature below $T_{\rm c}=12.5$~K a softening
of the lattice occurs. The softening corresponds to about 60~K decrease in the Debye temperature from its original value of 
$\sim 460$~K as determined by the isomer-shift vs. temperature data.

\section{ACKNOWLEDGMENTS}
Mr M. Lehtim\"aki and Mr T. Tynell are acknowledged for their help with performing the XRD and
Seebeck characterizations. Prof. P. Paturi of Wihuri research center is acknowledged for her contribution to the in-field Seebeck measurements. Mr S. Fr\"ojd\"o is acknowledged for his assistance with the sample synthesis. This work was partially supported by Tekes (No 1726/31/07) and Academy of Finland (No. 126528).

\bibliographystyle{model1a-num-names}

\end{document}